\documentclass[onecolumn]{aastex631}

\usepackage{multirow, graphicx, epsfig, color}
\DeclareGraphicsExtensions{.eps}

\shorttitle{Neutrino beaming patterns}
\shortauthors{B\"ottcher}

\begin{document}

\title{Beaming patterns of neutrino emission from photo-pion production in relativistic jets}

\author[0000-0002-8434-5692]{Markus B\"{o}ttcher}
\affil{Centre for Space Research, North-West University, Potchefstroom, 2531, South Africa}
\email{Markus.Bottcher@nwu.ac.za}

\begin{abstract}
In the light of growing evidence that blazars are responsible for part of the astrophysical 
very-high-energy neutrino flux detected by IceCube, models for neutrino production through 
photo-pion interactions in blazar jets have been developed. Evidence is also mounting that 
photon fields originating external to the jet are strongly favored over the co-moving primary
electron synchrotron photon field as target for photo-pion interactions. Even though those 
external photon fields appear highly anisotropic in the co-moving frame of the emission region, 
current models usually consider neutrino production to occur isotropically in the co-moving 
frame, resulting in a beaming pattern that is identical to intrinsically isotropic synchrotron 
and synchrotron self-Compton emission. In this paper, we derive the resulting beaming patterns 
of neutrinos produced by interactions wich external photon fields, taking into account all 
relevant anisotropy effects. It is shown that neutrino emission resulting from photo-pion 
production on a stationary and isotropic (in the AGN rest frame) external photon field is 
significantly more strongly beamed along the jet direction than intrinsically isotropic 
emission. For the most highly beamed sources, this implies that expected neutrino fluxes 
are grossly under-estimated or jet-power requirements for the production of a given neutrino
flux grossly over-estimated when not accounting for the proper Doppler boosting and beaming
characteristics. 
\end{abstract}

\section{\label{intro}Introduction}

Since the identification of a significant flux of astrophysical very-high-energy (VHE) neutrinos 
by the IceCube detector at the South Pole \citep{IceCube1,IceCube2}, evidence has been mounting 
that blazars, a class of jet dominated Active Galactic Nuclei (AGN) with their relativistic jets 
pointing close to our line of sight, are the sources of a fraction of this neutrino flux 
\citep[see, e.g.,][]{Garrappa19,Buson22,Plavin20,Plavin21,Plavin23}. While it has been demonstrated 
that blazars can not account for the entire astrophysical VHE neutrino flux \citep[see, e.g.,][]{Murase18}, 
the number of tentative associations between VHE neutrino events and flaring blazars is steadily increasing.
The highest-confidence identification so far, of TXS~0506+056 with the IceCube neutrino event IC-170922A 
\citep{IceCube3} as well as an excess of lower-energy neutrinos in 2014 -- 2015 \citep{IceCube4}, has been 
followed by several other, similarly tantalizing hints of correlations. Most notably, in 2021, the flaring 
blazar PKS 0735+17 has been found coincident not only with a single VHE neutrino event by IceCube, but 
potentially also with neutrino events detected by the Baikal-GVD experiment, the Baksan Underground 
Scintillation Telescope, and KM3NeT \citep[see][for details]{Sahakyan23}. 

Motivated by these developments, a large body of literature on theoretical modeling of the broadband 
electromagnetic radiation, from radio to $\gamma$-rays, and, simultaneously, neutrino emission from 
blazars has emerged in recent years \citep[for a recent review and relevant references, see, e.g.,][]{Boettcher19}. 
Due to the low number densities of particles expected in blazar jets, the dominant mode of neutrino 
production, assuming that protons or other nuclei can be accelerated to sufficiently high energies in 
the blazar jet environment, is through photo-pion production and subsequent pion and muon decay. In 
such interactions, each neutrino typically attains about 5~\% of the proton's energy. Due to the dominant 
$\Delta^+$ resonance at $E_{\Delta} = 1232$~MeV in the p$\gamma$ cross section, in combination with the 
typically steeply declining target photon number spectra, photo-pion interactions are strongly dominated
by the $\Delta^+$ resonance. These interactions are expected to take place in an emission region (referred 
to as ``blob'') that moves with a large bulk Lorentz factor $\Gamma \sim 10$ along the jet, resulting in 
relativistic Doppler boosting of photon and neutrino energies by a Doppler factor 
$\delta = \left( \Gamma \, [1 - \beta_{\Gamma} \cos\theta_{\rm obs}] \right)^{-1} \equiv 10 \delta_1$, 
where $\beta_{\Gamma}$ is the speed (normalized to the speed of light $c$) corresponding to the Lorentz factor 
$\Gamma$. The $\Delta^+$ resonance condition then requires for the production of neutrinos of energies 
$E_{\nu}^{\rm obs} = 10^{14} \, E_{14}$~eV, that the target photons have an energy of $E_t \sim 1.6 \, E_{14}^{-1} 
\, \delta_1$~keV in the co-moving frame of the emission region \citep[see, e.g.,][for details]{Gao19,Reimer19}. 
Photons of such energies produced co-spatially in the emission region are typically very scarce, leading to 
excessive requirements of jet power in relativistic protons in order to predict a measurable VHE neutrino flux 
\citep[e.g.,][]{Reimer19}. Thus, there is a growing consensus that external radiation fields, possibly 
approximately isotropic in the AGN rest frame, are energetically strongly preferred over co-moving radiation 
fields \citep[e.g.,][]{Righi19,Reimer19}. 

Radiation and neutrinos produced isotropically in the co-moving frame of the emission region, are Doppler 
boosted in frequency / energy by one Doppler factor $\delta$, such that $E^{\rm obs} = \delta \, E / (1 + z)$, 
where $z$ is the source's redshift. Throughout the text, we denote quantities in the observer's frame by a 
label ``obs", while unmarked quantities refer to the co-moving frame of the blob. Quantities in the AGN rest 
frame will be denoted by an asterisk (`$\ast$'). The Doppler boost in frequency/energy-integrated flux (between 
reference energies $E^{\rm obs}_1$ and $E^{\rm obs}_2$) is given by

\begin{equation}
F^{\rm obs} (\Omega_{\rm obs}) \equiv \int\limits_{E^{\rm obs}_1}^{E^{\rm obs}_2}  F^{\rm obs}_{E_{\rm obs}} 
(\Omega_{\rm obs}) \, d E^{\rm obs}
= {\delta^4 \over 4 \pi d_L^2} \int\limits_{E_1}^{E_2} L_E (\Omega) \, d E 
\label{Fdefinition}
\end{equation}

whereas the propagation directions of photons and neutrinos are related through $\mu = (\mu_{\rm obs} - 
\beta_{\Gamma})/(1 - \beta_{\Gamma} \mu_{\rm obs})$, with $\mu = \cos\theta$ and $\theta$ being the angle 
between the propagation direction and the jet axis. The above Doppler boosting pattern is typically assumed 
in most currently existing hadronic models for the multi-messenger emission from blazars. However, for the 
X-ray and $\gamma$-ray emission due to inverse Compton scattering of external radiation fields, it has been 
demonstrated \citep[e.g.,][]{Dermer95} that the beaming patterns of such radiation are quite different from 
the case of, e.g., isotropic synchrotron and synchrotron self-Compton (SSC) emission in the blob frame. A 
similar effect is expected for the case of photo-pion production on an external photon field. The expected
beaming patterns of secondaries in terms of $E L_{E}$ luminosities have, in fact, been derived in detail by 
\cite{Dermer12}, who focused on neutron production; however, the results are equally applicable to neutrino
production. They find a characteristic pattern of $\gamma_n \, L_n \propto \delta^5 \, {\gamma'}_p^2 \, N'_p
({\gamma'}_p)$, where $\gamma_n$ is the neutron's Lorentz factor and $N'_p ({\gamma'}_p)$ the number density of
parent protons in the co-moving frame. Nevertheless, even though this enhanced beaming has been pointed out
before, to this author's knowledge, the effect is routinely ignored in all existing codes for modeling 
multi-messenger emission from blazars. 

Therefore, in this manuscript, I evaluate the beaming patterns of the energy-integrated neutrino number flux 
(as measurable by IceCube and other neutrino detectors) produced in photo-pion production of co-moving and 
stationary external radiation fields, demonstrating numerically the drastic difference in those patterns. The 
paper is structured as follows: \S \ref{calculation} describes the formalism with which the neutrino flux 
as a function of viewing angle is evaluated. \S \ref{results} shows results for illustrative test cases of
power-law and thermal target radiation fields, and the findings are discussed in \S \ref{conclusions}.

\section{\label{calculation}Evaluation of beaming patterns}

The relevant quantity to estimate expected neutrino detection rates as a function of viewing angle with 
respect to the blazar jet, is the neutrino number flux, integrated over a given energy energy range, 
$E_{\nu, 1}^{\rm obs}$ to $E_{\nu, 2}^{\rm obs}$, which transforms as

\begin{equation}
\Phi_{\nu} (\Omega_{\rm obs}) \equiv \int\limits_{E_{\nu, 1}^{\rm obs}}^{E_{\nu, 2}^{\rm obs}} 
\Phi_{\nu, E^{\rm obs}} \, d \, E_{\nu}^{\rm obs} 
= {\delta^3 \, (1 + z) \over 4 \pi \, d_L^2} \, V_b \, \int\limits_{E_{\nu,1}}^{E_{\nu,2}} 
\dot n_{E_{\nu}} (\Omega) \, d E_{\nu},
\label{Phidefinition}
\end{equation}

where $V_b$ is the volume of the emission region and $d_L$ the luminosity distance to the source. 
For the evaluation of the differential production rate of neutrinos (per unit energy, per unit solid 
angle, per unit volume), $\dot n_{E_{\nu}} (\Omega)$, we employ a simple $\delta$ function approximation 
for the $p\gamma$ cross section, such that interactions are only considered in the $\Delta^+$ resonance, 
thus constraining the product of target photon energy $E_t$ and proton energy $E_p$ to be $E_t \, E_p \, 
(1 - \beta_p \, \chi) = E_{\Delta}^2$, where $\chi$ is the cosine of the angle between the propagation
directions of the proton and the target photon. We further constrain $E_p = 20 \, E_{\nu}$, as each 
neutrino typically takes up 5~\% of the original proton energy. The energy integral over the cross 
section is then approximated as $\Gamma_{\Delta} \, \bar{\sigma}$, with $\bar{\sigma} \approx 500 \, \mu$b 
and the width of the resonance, $\Gamma_{\Delta} \approx 115$~MeV. We further assume, for simplicity, 
that the relativistic proton energy distribution is isotropic in the blob frame. With these simplifications, 
the neutrino production rate reduces to

\begin{equation}
\dot n_{E_{\nu}} (\Omega) \approx c \, \bar{\sigma} \, \Gamma_{\Delta} \int\limits_{-1}^1 d\mu_t \, 
\int\limits_{0}^{2\pi} d\phi_t \, n_t (E_t, \Omega_t) \, n_p (E_p) \, (1 - \beta_p \, \chi),
\label{ndotgeneral}
\end{equation}

where the target photon energy is constrained to be $E_t = E_{\Delta}^2 / (20 \, E_{\nu} \, [1 - \beta_p \, \chi])$, 
and the target photon's propagation direction is given by $\hat{\Omega_t} = ( \sqrt{1 - \mu_t^2} \cos\phi_t, 
\sqrt{1 - \mu_t^2} \sin\phi_t, \mu_t)$. Due to the vastly dominant proton momentum in photo-pion production 
reactions, the neutrino propagation direction will be identical to the original proton propagation direction. 
Thus, $\hat{\Omega} = ( \sqrt{1 - \mu^2} \cos\phi, \sqrt{1 - \mu^2} \sin\phi, \mu)$ for both the proton and 
the neutrino. Hence, the interaction angle cosine is given by $\chi = \sqrt{1 - \mu^2} \, \sqrt{1 - \mu_t^2} 
(\cos\phi \, \cos\phi_t + \sin\phi \, \sin\phi_t) + \mu \, \mu_t$. 

For an illustrative example, we employ a simple power-law spectrum for the relativistic proton 
spectrum, such that

\begin{equation}
n_p (E_p) = n_{p,0} \left( {E_p \over E_{p,0}} \right)^{-p} \;\;\;\;\;\;\;\; {\rm for} \;\; E_{p,1} \le E_p \le E_{p,2};
\label{p_powerlaw}
\end{equation}
where $E_{p,0}$ is an arbitrary reference energy and $n_{p,0}$ is a normalization constant with units of 
cm$^{-3}$~eV$^{-1}$. 

For the target photon fields, we will consider the cases of simple power-law fields (as, e.g., the co-moving 
electron-synchrotron field or the X-ray spectrum of an accretion-disk corona) and (quasi-)thermal photon fields
\citep[such as, e.g., the dust torus or the broad-line-region, whose spectrum can, for the purpose of calculating 
external-Compton or neutrino emission, be well approximated by a blackbody spectrum, see, e.g.,][]{Boettcher13}. 
For both spectral shapes, the case of a co-moving radiation field, isotropic in the blob frame, will be compared
to a photon field that is isotropic in the AGN rest frame.

\subsection{\label{powerlaws}Power-law target photon spectra}

The spectrum of a power-law target photon field is represented as 

\begin{equation}
n_t (E_t) = n_{t,0} \left( {E_t \over E_{t,0}} \right)^{-\alpha} \;\;\;\;\;\;\;\; {\rm for} \;\; E_{t,1} \le E_t \le E_{t,2}
\label{ph_powerlaw}
\end{equation}
where $E_{t,0}$ is an arbitrary reference energy and $n_{t,0}$ is a normalization constants as for the proton
spectrum.

\subsubsection{\label{pl_comoving}Isotropic co-moving power-law target photon field}

In the case of an isotropic target photon field in the co-moving frame, we may set $n_t (E_p, \Omega) = n_t (E_p) 
/ (4 \pi)$ in Eq. \ref{ndotgeneral}.
A simple analytic estimate for the resulting beaming pattern may be found if the conditions for $E_t$ and $E_p$ can be met within the limits set in 
Eq. \ref{p_powerlaw} for the entire neutrino energy range from $E_{\nu, 1}$ to $E_{\nu, 2}$. The complete angle integration in Eq. \ref{ndotgeneral} can only be
solved numerically, and results are shown in \S \ref{results}. For an analytic estimate, we now consider only head-on collisions, in which case
the reactions are most efficient, i.e., we set $\chi = -1$. Furthermore, as the protons need to be highly relativistic, we set $\beta_p = 1$. With these
simplifications, we find that the production rate will scale as

\begin{equation}
\dot{n}_{E_{\nu}} (\Omega) \propto \left( {20 \, E_{\nu} \over E_{p,0}} \right)^{-p} \, \left( {E_{\Delta}^2 \over 40 \, E_{\nu} \, E_{t,0}} \right)^{-\alpha}
\propto E_{\nu}^{\alpha - p}. 
\label{ndotiso}
\end{equation}
  
Plugging this into Eq. (\ref{Phidefinition}) finally yields the approximate beaming pattern as 

\begin{equation}
\Phi_{\rm obs} (\Omega_{\rm obs}) \propto \delta^{2 - \alpha + p} \, (1 + z)^{2 + \alpha - p}.
\label{com_approx}
\end{equation}

This will be compared to a detailed numerical calculation in \S \ref{results}.

\subsubsection{\label{pl_stationary}Stationary power-law target photon field}

We now consider a target photon field that is stationary and isotropic in the AGN rest frame. In this case, 
we may write $n_t^{\ast} (E_t^{\ast}, \Omega^{\ast}) = n_t^{\ast} (E_t^{\ast}) / (4 \pi)$, which is transformed 
into the blob frame as 

\begin{equation}
n_t (E_t, \Omega_t) = {n_t^{\ast} (E_t^{\ast}) \over 4 \pi \, \delta_{\ast}^2},
\label{ntransform}
\end{equation}
where the boosting from the AGN to the blob frame is characterized by $\delta_{\ast} = 
\Gamma \, (1 + \beta_{\Gamma} \, \mu_t)$. Inserting Eq. \ref{ntransform} into Eq. \ref{ndotgeneral} and Eq. 
\ref{Phidefinition} provides an integral expression for the angle-dependent neutrino flux. Results of numerically 
evaluating those integrals will be presented in \S \ref{results}.

Also in this case, we may derive a simple estimate of the resulting beaming pattern assuming that the energy 
conditions can be met throughout the considered spectral range. For highly relativistic bulk Motion, 
$\Gamma \gg 1$, the beaming of the radiation into the comoving blob frame may be approximated by setting 
$\mu_t = -1$, i.e., assuming all target photons enter directly from the front. In this case, the Doppler factor
$\delta_{\ast} = \Gamma (1 - \beta_{\Gamma}) \approx 1/(2 \Gamma)$. Hence, $E_t \approx 2 \, \Gamma \, E_t^{\ast}$, 
and from the transformation $u_t = \int n_t (E_t) \, E_t \, d E_t \sim \Gamma^2 \, u_t^{\ast}$ it follows that 
$n_t (E_t) \approx n_t^{\ast} (E_t^{\ast})$, while $\chi = - \mu$ and $\beta_p \approx 1$. With these simplifications, 
we find

\begin{equation}
\dot{n}_{E_{\nu}} (\Omega) \propto \, \left( {20 \, E_{\nu} \over E_{p,0}} \right)^{-p} \, 
\left( {E_{\Delta}^2 \over 40 \, \Gamma \, E_{\nu} \, [1 + \mu]} \right)^{-\alpha}.
\label{ndotforward}
\end{equation}

Plugging this into Eq. \ref{Phidefinition}, noting that $(1 + \mu) \approx \delta \, (1 + \mu_{\rm obs}) 
/ (2 \Gamma)$, yields 

\begin{equation}
\Phi_{\rm obs} (\Omega_{\rm obs}) \propto \delta^{3 + p} \, (1 + \mu_{\rm obs})^{\alpha} \, (1 + z)^{2 + \alpha - p}. 
\label{stat_approx}
\end{equation}

This indicates a more pronounced Doppler beaming than in the co-moving isotropic case. This 
approximation will be compared to full numerical solutions in the following section.

\subsection{\label{thermal}Thermal target photon fields}

The thermal radiation-field case is represented as a blackbody spectrum with

\begin{equation}
n_t (E_t) = n_{t,0} { E_t^2 \over exp(E_t / k_B T) - 1} 
\label{ph_blackbody}
\end{equation}
where $k_B$ is the Boltzman constant and $T$ the temperature.

\subsubsection{\label{th_comoving}Co-moving isotropic thermal photon field}

As for the case of the power-law photon spectra, Eq. \ref{ndotgeneral} may be solved numerically, 
setting $n_t (E_p, \Omega) = n_t (E_p) / (4 \pi)$.

An analytical estimate may be derived by approximating the photon field as monochromatic at an
energy $E_t = 2.8 \, k_B T$. This fixes the proton energy for interactions in the $\Delta$ resonance
to $E_p = E_{\Delta}^2 / (2 \, E_t)$ and, consequently, the neutrino energy to $E_{\nu} = E_{\Delta}^2
/ (40 \, E_t)$. This corresponds to an observed neutrino energy of 

\begin{equation}
E_{\nu}^{\rm obs} = {\delta \over 1 + z} \, E_{\nu} \approx 1.6 \, {\delta_1 \over 1 + z} \, T_6^{-1}
\, {\rm PeV}
\label{Enu_th_comoving}
\end{equation}
where $T_6 = T / (10^6 \, {\rm K})$.

Thus, co-moving target photon temperatures in excess of $10^6$~K are required in order to produce 
neutrinos in the sub-PeV energy range. No such target photon fields are plausibly produced within
the emission region, but this can be thought of an external, thermal radiation field, Doppler boosted
into the emission region by a factor $\sim 2 \, \Gamma$, i.e., $T \sim 2 \, \Gamma \, T^{\ast}$. In 
the currently generally adopted framework of treating neutrino production as isotropic in the blob 
rest frame, this would then lead to a beaming charactristic of 

\begin{equation}
\Phi_{\rm obs} (\Omega_{\rm obs}) \propto \delta^3 \, (1 + z) 
\label{th_comoving_approx}
\end{equation}
if the neutrino energy as given by Eq. \ref{Enu_th_comoving} is within the observable energy range.

\subsubsection{\label{th_stationary}Stationary isotropic thermal photon field}

The proper beaming characteristics in this case are as described in \S \ref{pl_stationary}. A monochromatic 
approximation, as in the previous sub-section, then fixes the proton energy to $E_p = E_{\Delta}^2 / (\delta \,
[1 + \mu_{\rm obs}] \, E^{\ast}_t)$ and the observed neutrino energy to 

\begin{equation}
E_{\nu}^{\rm obs} \approx {\delta \over 1 + z} \, {E_{\Delta}^2 \over 112 \, (1 + \mu) \, \Gamma \, k_B \, T^{\ast}}
\approx {320 \, {\rm TeV} \over (1 + z) \, T^{\ast}_6 \, (1 + \mu_{\rm obs})} 
\label{Enu_th_stationary}
\end{equation}
Remarkably, the Doppler and bulk Lorentz factor dependence cancels out in this simple derivation, leaving only
a moderate viewing-angle dependence. For a jet viewed directly head-on, $\mu_{\rm obs} = 1$, and the neutrino 
energy is reduced by one factor of $\delta$ compared to the stationary-photon-field case, as expected, when 
comparing identical temperatures $T$ and $T^{\ast}$. 

The expected neutrino beaming pattern then reduces to

\begin{equation} 
\Phi_{\rm obs} (\Omega_{\rm obs}) \propto \delta^{3 + p} \, (1 + \mu_{\rm obs})^p \, (1 + z) 
\label{th_stationary_approx}
\end{equation}
similar to Eq. \ref{stat_approx} for a power-law target photon field.

\section{\label{results}Results}

We now consider simple, illustrative test cases for a full numerical solution to 
Eqs. \ref{Phidefinition} and \ref{ndotgeneral} to evaluate the neutrino flux in the range
$E_1^{\rm obs} = 10$~TeV to $E_2^{\rm obs} = 1$~ PeV as a function of viewing angle. 
We employ a power-law proton distribution spanning from $E_{p, 1} = 1$~GeV to 
$E_{p,2} = 100$~PeV and explore different proton spectral indices between 2 and 3, various 
power-law spectral indices and radiation temperatures for the target photon fields,
and bulk Lorentz factors ranging from $\Gamma = 5$ to 20.

\subsection{\label{pl_results}Power-law target photon spectra}

The target photon fields in the case of power-law spectra are chosen between $E_{t,1} = 1$~eV 
and $E_{t,2} = 100$~keV, and we explore spectral indices between 2 and 3. 

\begin{figure}
\includegraphics[width=8.5cm]{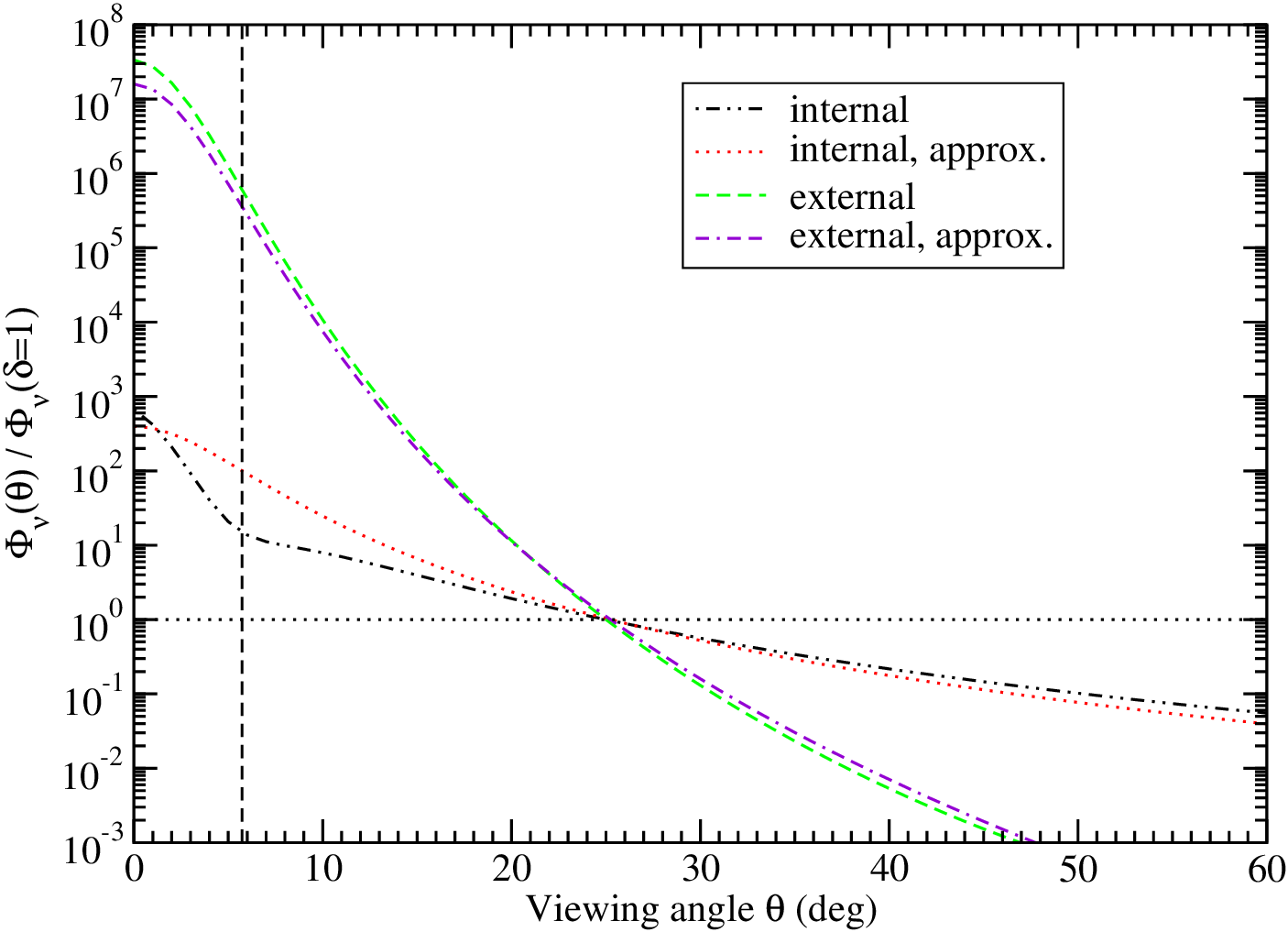} \qquad
\includegraphics[width=8.5cm]{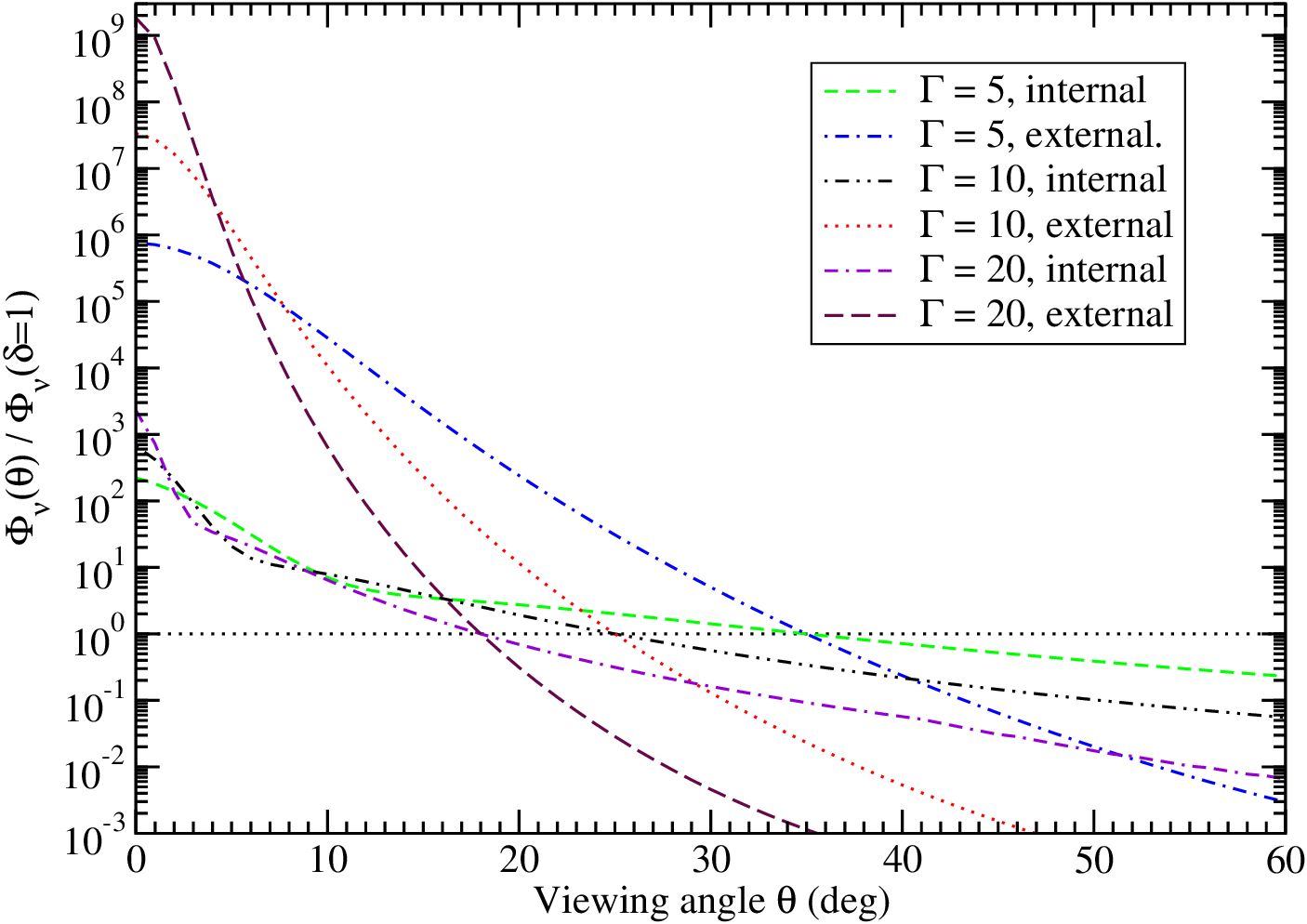}
\caption{\label{alpha25_p25} {\it Left:} Neutrino beaming patterns for default parameters, 
$p = \alpha = 2.5$, and $\Gamma = 10$, comparing the cases of isotropic external
and co-moving radiation fields, and comparing the full numerical calculation to 
the analytical approximations (Eqs. \ref{com_approx} and \ref{stat_approx}), 
as indicated by the legends. The vertical dashed line
indicates the superluminal angle of $\theta_{\rm obs} = 1/\Gamma$. 
{\it Right:} Neutrino beaming patterns for the same parameters as in the left panel,
but varying $\Gamma$ between 5 and 20. }
\end{figure}

To illustrate the beaming patterns, we normalize each neutrino-flux vs. viewing-angle
curve to the unbeamed flux value, i.e., the value at an angle where $\delta = 1$, which 
occurs for $\mu_{\rm obs} = \Gamma \, \beta_{\Gamma} / (\Gamma + 1)$. The left panel of
Fig. \ref{alpha25_p25} 
shows the comparison of the neutrino beaming patterns for the cases of a co-moving and a 
stationary external photon field for $\Gamma = 10$ and $p = \alpha = 2.5$ and contrasts 
them with the analytical approximations of Eqs. \ref{com_approx} and \ref{stat_approx}.
It is immediately clear that the Doppler enhancement at small viewing angle is stronger
by several orders of magnitude in the case of the external radiation field, compared to
a comoving one, as expected. Also the angular pattern appears significantly more strongly
peaked in the external field case. Especially in the case of the internal target photon
field, the analytical approximation derived in Eq. \ref{com_approx} only gives a rough
guide to the beaming pattern, but deviates from the full numerical calculation by about
an order of magnitude for viewing angles near $1/\Gamma$. In the following parameter study,
we will therefore only plot the full numerical results. 

The right panel of Fig. \ref{alpha25_p25} illustrates the dependence of the neutrino
beaming patterns on the bulk Lorentz factor. As expected, especially in the case of a
stationary external radiation field, the beaming becomes narrower and stronger at
small angles, in rough agreement with the estimated $\delta^{5.5}$ scaling
of Eq. \ref{stat_approx}. This effect is significantly less pronounced in the 
case of a co-moving target photon field, also in agreement with the estimated 
$\delta^2$ scaling of Eq. \ref{com_approx}. 

\begin{figure}
\includegraphics[width=8.5cm]{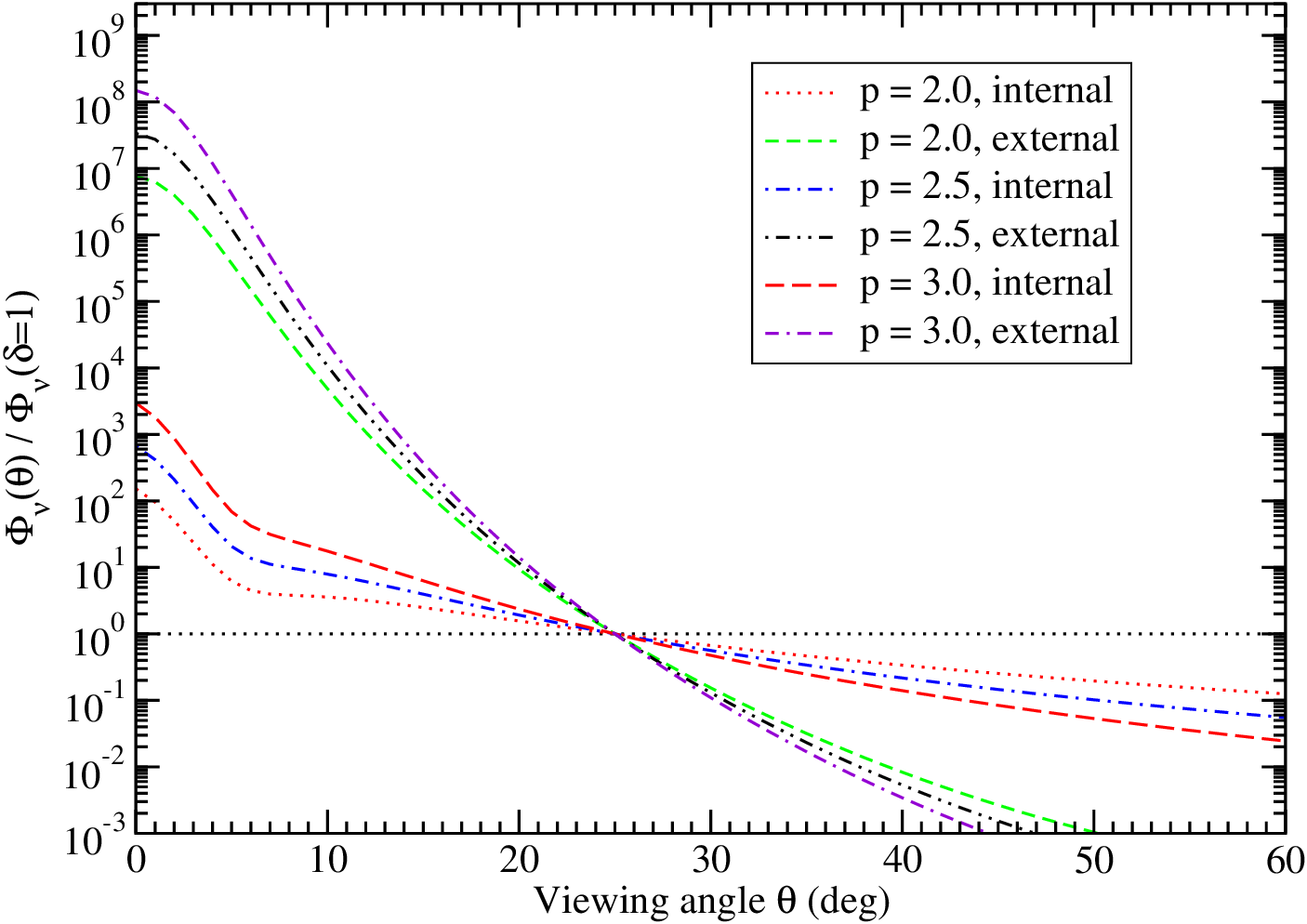} \qquad
\includegraphics[width=8.5cm]{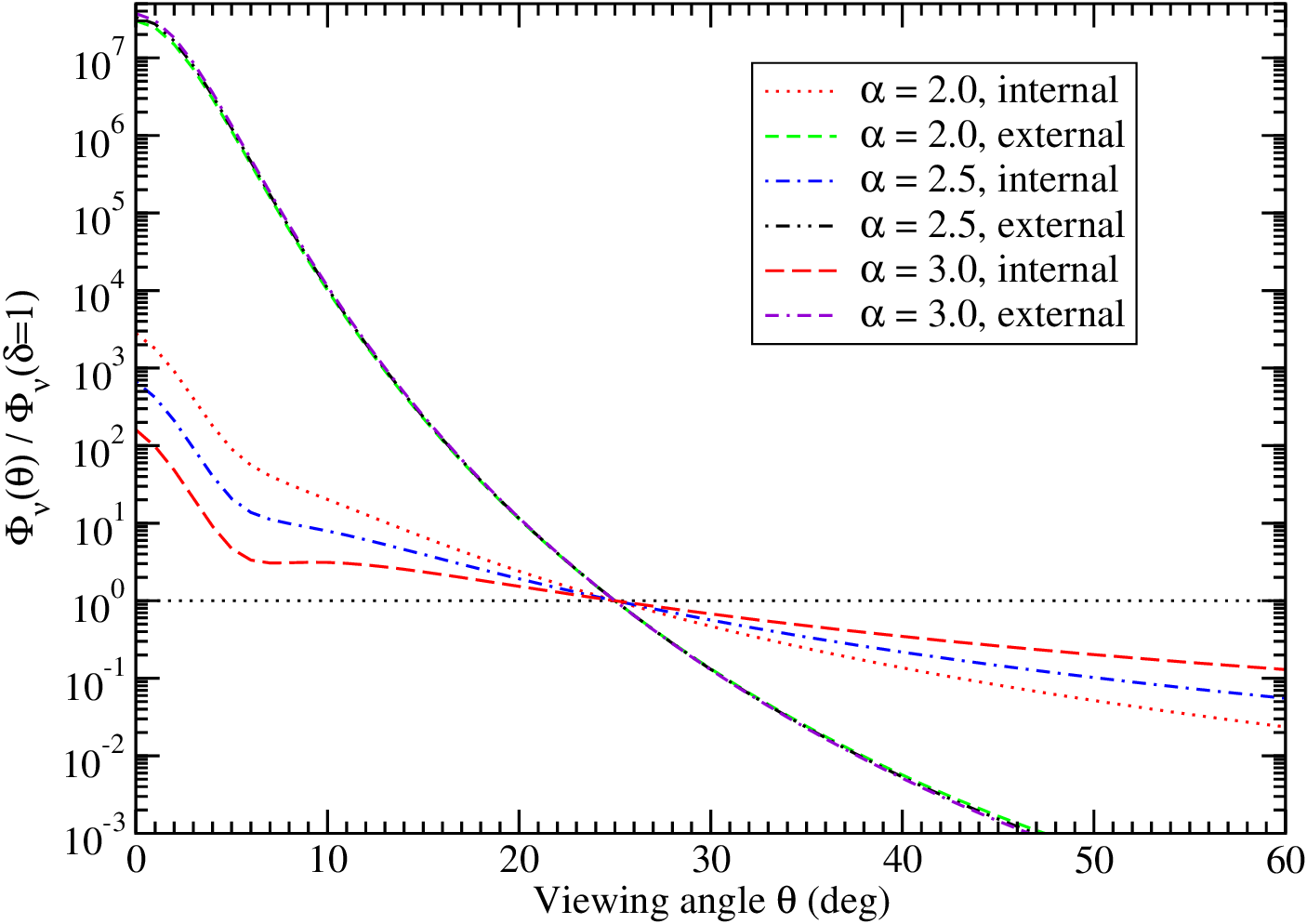}
\caption{\label{alphapvar} {\it Left:} Effect of the proton spectral index on the
neutrino beaming patterns, for $\Gamma = 10$, and $\alpha = 2.5$ and three different 
values of $p$. 
{\it Right:} Effect of the target photon spectral index on the neutrino beaming patterns,
for $\Gamma = 10$, $p = 2.5$, and four different values of $\alpha$}
\end{figure}

Fig. \ref{alphapvar} illustrates the effect of a changing proton (left) and target photon 
(right) spectral index. In agreement with the analytical estimates (Eq. \ref{com_approx}
and \ref{stat_approx}), there is a moderate $\sim \delta^p$ dependence on the proton 
spectral index in both cases of the target photon field. The Doppler boosting in the case 
of a co-moving target photon field shows the expected $\delta^{-\alpha}$ dependence on
the target-photon spectral index, while the very weak $\alpha$ dependence expected from
Eq. \ref{stat_approx} is seen for the case of the stationary target photon field.

\subsection{\label{th_results}Thermal target photon fields}

As mentioned in Section \ref{thermal}, the $\Delta^+$ resonance condition, in combination 
with the narrow spectrum of a thermal blackbody radiation field, set stringend constraints 
on the target photon blackbody temperature for the production of neutrinos in the IceCube 
energy range. Co-moving temperatures in excess of $\sim 10^6$~K are required, corresponding
to temperatures in the stationary frame of $\sim 10^5$~K. 

\begin{figure}
\includegraphics[width=8.5cm]{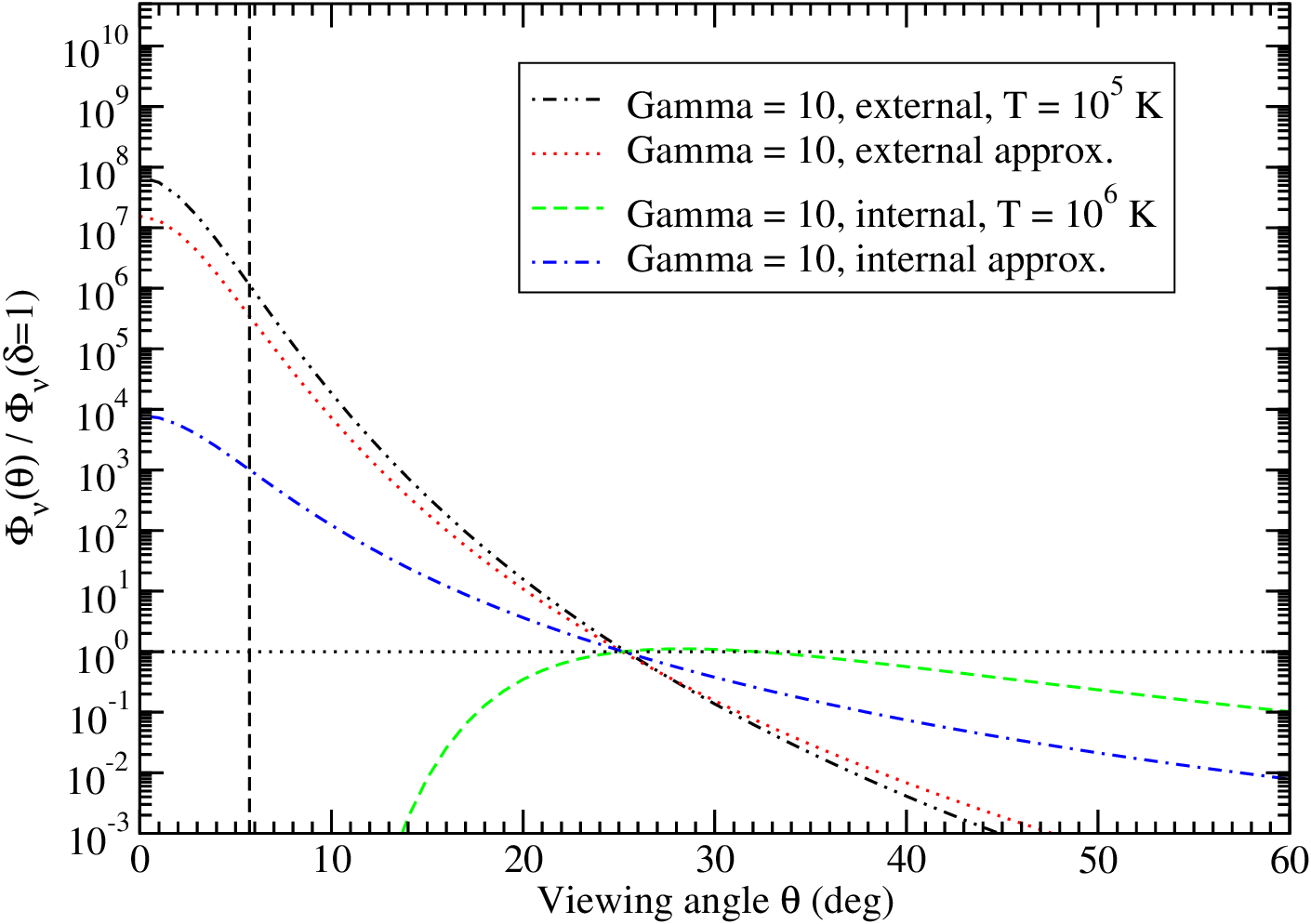} \qquad
\includegraphics[width=8.5cm]{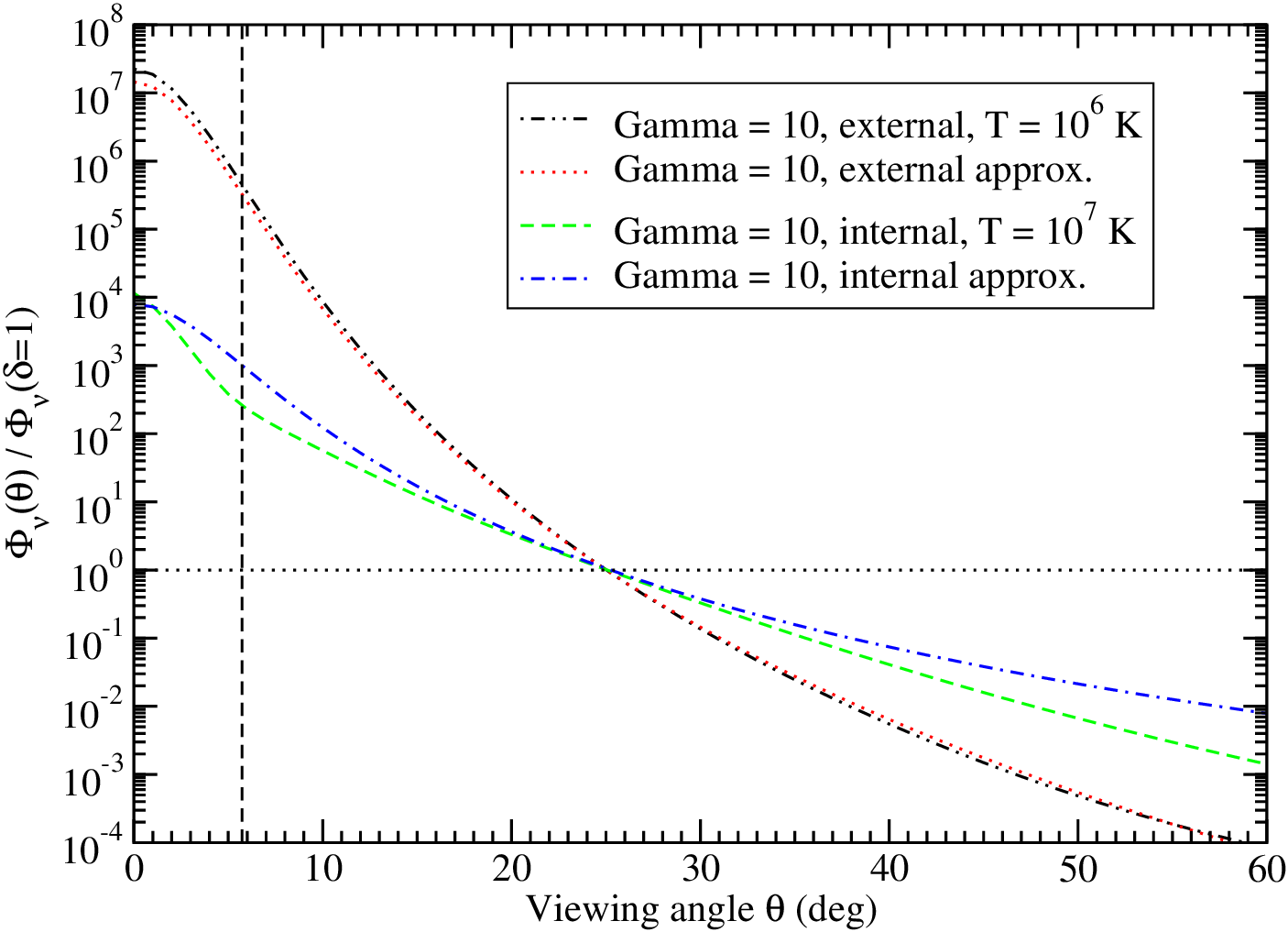}
\caption{\label{Fig:Thermal}Beaming patterns for neutrino production on thermal target photon fields 
of (black dot-dot-dashed and green dashed), compared to the monochromatic approximation (red dashed
and blue dot-dashed) under the assumption that neutrinos in the 10~TeV -- 1~PeV range can be produced 
in the $\Delta^+$ resonance. Left panel: $T^{\ast} = 10^5$~K in the stationary frame, compared to 
$T = 10^6$~K in the co-moving frame. Right panel: $T^{\ast} = 10^6$~K in the stationary frame,
compared to $T = 10^7$~K in the co-moving frame. }
\end{figure}

Fig. \ref{Fig:Thermal} compares the beaming patterns for stationary and co-moving thermal radiation
fields, chosing $T = \Gamma \, T^{\ast}$ in order to compare radiation fields with comparable co-moving
blackbody temperatures. Statinary temperatures of $T^{\ast} = 10^5$~K and $T^{\ast} = 10^6$~K are 
considered. The left panel of the figure shows that in the $T = 10^6$~K in the co-moving frame, for 
a small viewing angle, the $\Delta^+$ resonance condition can hardly be met by any thermal target 
photons to produce neutrinos in the IceCube energy range, thus suppressing neutrino emission. For 
the higher-temperature case (right panel) the numerical results agree well with the mono-chromatic
approximation, and the difference in beaming intensity in the forward direction is, as in the case
of power-law target photon fields, a factor of several thousands stronger for photo-pion production 
on external photon fields than for co-moving photon fields, for a characteristic bulk Lorentz factor
of $\Gamma = 10$. Other dependencies on $p$ and $\Gamma$ are similar as for the power-law target 
photon fields discussed in the previous sub-section and not repeated here for the sake of brevity.

\section{\label{conclusions}Discussion and conclusions}

We have derived the dependence of the Doppler boosting and beaming of neutrino emission
from a blazar jet on the viewing angle and, thus, the Doppler factor $\delta$. In particular,
we contrasted these beaming patterns between the cases of a co-moving target photon field,
which is isotropic in the co-moving frame of the emission region, and an external target
photon field, which is isotropic in the stationary AGN frame. Simple illustrative test
cases of straight power-law proton spectra were considered, with both power-law and blackbody
target photon spectra. Power-law target photon spectra in the X-ray regime are plausibly
provided by a hot corona associated with the accretion flow. The very high temperatures 
required for the thermal radiation fields could originate in a very hot, near-Eddington 
accretion disk. 

We found that the Doppler boosting and beaming has a significantly stronger dependence on the 
Lorentz factor and viewing angle in the case of a stationary external target photon field,
compared to the case of a co-moving target photon field. A rough analytical estimate of 
the beaming pattern for an external target photon field of $\Phi_{\rm obs} \propto \delta^{3+p}
\, (1 + \mu_{\rm obs})^\alpha$ is approximately reproduced by detailed numerical 
calculations. As it has been found in many studies that external target photon fields are 
favoured for neutrino production in blazar jets, our results suggest that the most 
favoured candidate neutrino-emitting blazars are those that have their jets most closely 
aligned with the line of sight to Earth. This seems to be supported by the fact that out 
of the three blazars with the highest Doppler factors ($\delta > 100$) found in the MOJAVE 
sample of 309 blazar jets analyzed by \cite{Homan21}, two are amongst the tentative neutrino 
counterparts identified by \cite{Plavin23}. Most importantly, for characteristic bulk Lorentz
factors of $\Gamma \sim 10$, neutrino emission from photo-pion interactions on external radiation
fields in the forward direction is boosted by factors of several thousands more strongly than 
in the current standard assumption of isotropic neutrino production in the co-moving frame. 
For the most highly beamed sources, this means that the measurable neutrino flux may be 
under-estimated by factors of thousands or, conversely, the proton power requirements to 
produce a given neutrino flux, may be over-estimated by factors of thousands. 

A limitation of this work is the restriction to interactions in the $\Delta^+$ resonance.
This is a very good approximation in the case of broad power-law target photon spectra
\citep[see, e.g.,][]{Muecke99}. However, in the case of narrow, (quasi-)thermal target 
photon fields, higher center-of-momentum interactions, including multi-pion production
channels, are likely to play a much more significant role \citep[see, e.g.,][]{Fiorillo21}. 
A detailed study including these channels will be the subject of future work.

\begin{acknowledgments}
The author thanks Yuri Kovalev and Alexander Plavin for stimulating discussions which
motivated this work, as well as Foteini Oikonomou, Haocheng Zhang, and Matteo Cerruti
for careful reading of this manuscript and constructive comments. 
\end{acknowledgments}

\end{document}